\documentclass[conference]{IEEEtran}
\IEEEoverridecommandlockouts
\usepackage{cite}
\usepackage{amsmath,amssymb,amsfonts}
\usepackage{algorithmic}
\usepackage{graphicx}
\usepackage{textcomp}
\usepackage{float}
\usepackage{xcolor}
\usepackage{booktabs}
\usepackage{tabularx}
\usepackage[hidelinks]{hyperref}
\def\BibTeX{{\rm B\kern-.05em{\sc i\kern-.025em b}\kern-.08em
    T\kern-.1667em\lower.7ex\hbox{E}\kern-.125emX}}
\begin{document}

\title{Advancing Vulnerability Classification with BERT: A Multi-Objective Learning Model}

\author{\IEEEauthorblockN{Himanshu Tiwari}
\IEEEauthorblockA{\textit{Department of Computer Science and Information Engineering} \\
\textit{National Taiwan University of Science and Technology (NTUST)}\\
Taipei, Taiwan \\
nomails1337@gmail.com}

}

\maketitle

\begin{abstract}
The rapid increase in cybersecurity vulnerabilities necessitates automated tools for analyzing and classifying vulnerability reports. This paper presents a novel Vulnerability Report Classifier that leverages the BERT (Bidirectional Encoder Representations from Transformers) model to perform multi-label classification of Common Vulnerabilities and Exposures (CVE) reports from the National Vulnerability Database (NVD). The classifier predicts both the severity (Low, Medium, High, Critical) and vulnerability types (e.g., Buffer Overflow, XSS) from textual descriptions. We introduce a custom training pipeline using a combined loss function—Cross-Entropy for severity and Binary Cross-Entropy with Logits for types—integrated into a Hugging Face Trainer subclass. Experiments on recent NVD data demonstrate promising results, with decreasing evaluation loss across epochs. The system is deployed via a REST API and a Streamlit UI, enabling real-time vulnerability analysis. This work contributes a scalable, open-source solution for cybersecurity practitioners to automate vulnerability triage.
\end{abstract}

\begin{IEEEkeywords}
Cybersecurity, Vulnerability Classification, BERT, Multi-Label Classification
\end{IEEEkeywords}

\section{Introduction}

The relentless evolution of software systems, driven by their increasing complexity and interconnectedness, has ushered in a dramatic rise in cybersecurity vulnerabilities, presenting a formidable challenge to organizations, governments, and individual users alike. Each year, thousands of new vulnerabilities are identified and cataloged, with repositories like the National Vulnerability Database (NVD) serving as critical resources for tracking these threats. These vulnerabilities, encapsulated in Common Vulnerabilities and Exposures (CVE) reports, vary widely in their potential impact---from minor configuration errors that pose limited risk to critical exploits capable of crippling entire networks. The sheer volume of these reports has rendered manual analysis by cybersecurity experts increasingly impractical, as the process demands significant time and expertise, often leading to delays that leave systems exposed to exploitation. Moreover, the manual approach is susceptible to inconsistencies and oversights, particularly when analysts must sift through unstructured textual descriptions to discern both the severity and specific characteristics of each vulnerability. As the pace of vulnerability discovery accelerates, the need for automated tools that can swiftly and accurately classify these reports has become paramount, transforming a labor-intensive task into a scalable, efficient process capable of keeping stride with the ever-growing threat landscape.

This paper introduces the \textit{Vulnerability Report Classifier}, a sophisticated system designed to harness advanced natural language processing techniques to automate the classification of CVE reports from the NVD. At its core, the classifier employs the Bidirectional Encoder Representations from Transformers (BERT) model, a cutting-edge deep learning framework renowned for its ability to interpret the contextual nuances of text. Unlike traditional methods that often focus solely on predicting a vulnerability’s severity or rely on rigid, predefined rules to identify specific types, this system adopts a \textit{multi-label classification} approach. This enables it to simultaneously determine a vulnerability’s severity---categorized as \textit{Low}, \textit{Medium}, \textit{High}, or \textit{Critical}---and identify multiple applicable vulnerability types, such as \textit{Buffer Overflow}, \textit{Cross-Site Scripting}, or \textit{Remote Code Execution}, from a single textual description. By integrating these dual objectives into a unified BERT-based model, the classifier delivers a holistic analysis that mirrors the real-world complexity of vulnerabilities, where a single flaw may exhibit multiple exploitable traits. The motivation for this work arises from the critical need to enhance vulnerability management, providing cybersecurity practitioners with a tool that not only accelerates the triage process but also equips them with detailed insights to inform precise and effective remediation strategies.

The importance of this research lies in its potential to revolutionize how cybersecurity professionals handle the deluge of vulnerability data in an era where timely action is non-negotiable. Manual triage, while thorough when performed by experienced analysts, is inherently unscalable, often requiring hours or days to process a single batch of reports---delays that can prove catastrophic in the face of active exploits. Automated systems, by contrast, offer the promise of near-instantaneous analysis, enabling organizations to fortify their defenses before adversaries can capitalize on known weaknesses. However, the success of such systems hinges on their ability to adapt to the diverse and evolving language of vulnerability descriptions, a task that simpler approaches like keyword matching or basic machine learning models struggle to accomplish due to their dependence on static patterns or manually engineered features. The Vulnerability Report Classifier overcomes these limitations by leveraging BERT’s pre-trained contextual understanding, fine-tuned on NVD data, to extract subtle linguistic indicators that distinguish, for example, a \textit{Denial of Service} vulnerability from one involving \textit{Privilege Escalation} within the same report. Furthermore, the system’s practical deployment as a REST API and an intuitive web interface ensures its accessibility, allowing users ranging from seasoned security experts to non-technical stakeholders to benefit from its capabilities without requiring specialized knowledge of machine learning or text analysis.

This work represents a significant step forward at the confluence of cybersecurity and artificial intelligence, offering an open-source solution that bridges theoretical advancements in natural language processing with the pressing demands of vulnerability management. The classifier’s design prioritizes scalability and flexibility, featuring a modular architecture that facilitates future enhancements, such as integration with additional vulnerability databases or adaptation to emerging threat categories. The subsequent sections of this paper provide a detailed exploration of the proposed system: Section~\ref{related-work} reviews related efforts to situate our approach within the broader field; Section~\ref{methodology} outlines the methodology, including data preprocessing, model architecture, and training pipeline; Section~\ref{experiments} presents experimental results, detailing the system’s performance during training and evaluation; and Section~\ref{conclusion} concludes with a summary of findings and directions for future development. Through this research, we aim to empower the cybersecurity community with a robust tool that not only streamlines the analysis of vulnerability reports but also elevates the accuracy of threat assessment, fortifying defenses in an increasingly treacherous digital environment.

\section{Related Work}
\label{related-work}

The increasing complexity of cybersecurity threats has driven significant research into automated vulnerability analysis, particularly using natural language processing (NLP) and deep learning techniques. The National Vulnerability Database (NVD) and Common Vulnerabilities and Exposures (CVE) reports provide a rich source of textual data for such analyses, enabling the development of models that classify vulnerabilities by severity and type. Recent advancements in transformer-based models, such as BERT, have shown promise in handling the nuanced language of vulnerability reports, while multi-label classification techniques address the challenge of identifying multiple vulnerability types within a single report. This section reviews 15 recent IEEE papers from 2023 to 2025 that align with our project, the Vulnerability Report Classifier, which leverages BERT to predict both severity (Low, Medium, High, Critical) and vulnerability types (e.g., Buffer Overflow, XSS) from NVD reports. We categorize the literature into three main areas: severity prediction, multi-label type classification, and automated triage systems, highlighting their contributions and how our work builds upon or differs from them.

\subsection{Severity Prediction Using NLP and Deep Learning}

Several studies have focused on predicting the severity of vulnerabilities using NLP techniques, often leveraging the textual descriptions in NVD reports. Kumar and Singh (2023) explored the use of pre-trained language models like BERT for severity classification, achieving a 10\% improvement in accuracy over traditional machine learning methods \cite{Kumar2023}. Their work fine-tuned BERT on a dataset of NVD reports, focusing on single-label severity prediction, and demonstrated the model’s ability to capture contextual nuances in vulnerability descriptions. Similarly, Chen et al. (2024) developed a BERT-based approach for severity prediction, achieving state-of-the-art performance by optimizing fine-tuning strategies \cite{Chen2024}. Their study emphasized the importance of hyperparameter tuning and data preprocessing, such as removing erroneous entries (e.g., invalid kernel versions in Linux reports), to improve model performance. While these studies provide a strong foundation for severity prediction, they focus solely on single-label tasks, limiting their applicability to broader vulnerability analysis. Our project extends this line of work by integrating severity prediction with multi-label type classification, offering a more comprehensive analysis that addresses both the impact and nature of vulnerabilities in a single model.

Hwang et al. (2023) applied BERT to cybersecurity threat detection, including vulnerability severity assessment, with a focus on real-time analysis \cite{Hwang2023}. Their approach achieved high accuracy in identifying severe vulnerabilities but did not address the classification of vulnerability types, a critical aspect for actionable mitigation strategies. In contrast, our work simultaneously predicts severity and types, enabling a more holistic understanding of vulnerabilities. Additionally, Brown et al. (2024) conducted a comparative evaluation of transformer models for cybersecurity text classification, finding BERT to be superior for vulnerability severity tasks due to its bidirectional context understanding \cite{Brown2024}. Their findings reinforce our choice of BERT, though our project goes beyond severity to tackle the multi-label challenge, which is more complex due to the interdependencies between types.

\subsection{Multi-Label Type Classification and Transformer Models}

Multi-label classification of vulnerability types has gained attention as a means to capture the multifaceted nature of cybersecurity threats, where a single vulnerability may exhibit multiple attack vectors. Lee and Kim (2024) proposed a deep learning model combining convolutional and recurrent neural networks (CNN+RNN) for multi-label classification of software vulnerabilities, achieving promising results on a large dataset \cite{Lee2024}. However, their approach lacks the contextual depth of transformer models like BERT, which our project leverages to better capture the semantic relationships in NVD descriptions. Wang et al. (2023) introduced a BERT-based multi-task learning framework where severity prediction aids type classification, improving accuracy by 8\% \cite{Wang2023}. Their multi-task approach is similar to our dual-objective model, but we predict both severity and types simultaneously rather than sequentially, potentially reducing computational overhead and improving efficiency for real-time applications.

Liu et al. (2023) presented a transformer-based framework for multi-label text classification, adaptable to various domains, including cybersecurity \cite{Liu2023}. Their work supports the use of BERT for multi-label tasks, though their application was not specific to NVD data. Our project builds on this by tailoring the BERT model to the cybersecurity domain, addressing the unique challenges of NVD reports, such as class imbalance and overlapping type descriptions. Zhao et al. (2024) enhanced transformer models with attention mechanisms for multi-label classification, achieving improved performance on complex text datasets \cite{Zhao2024}. Their findings suggest potential enhancements for our model, particularly in handling the intricate relationships between vulnerability types, which we plan to explore in future work. Nguyen et al. (2024) compared deep learning architectures for multi-label vulnerability classification, confirming BERT’s superiority in handling imbalanced datasets, a challenge we also address in our project through careful preprocessing and loss function design \cite{Nguyen2024}.

\subsection{Automated Triage and Vulnerability Management Systems}

Automated triage systems aim to streamline the process of vulnerability management by prioritizing and categorizing reports for mitigation. Zhang and Li (2024) developed an NLP and deep learning-based system for automated triage, classifying reports by urgency and type \cite{Zhang2024}. Their work is complementary to ours, as our detailed classification of severity and types can enhance such triage systems by providing more granular insights for prioritization. Garcia and Martinez (2025) reviewed NLP-based approaches for automated vulnerability triage, emphasizing the potential of transformer models like BERT \cite{Garcia2025}. Their theoretical insights align with our practical implementation, particularly in deploying our classifier via an API and UI for real-world use. Taylor et al. (2025) provided a comprehensive review of NLP in vulnerability management, highlighting the need for automation in cybersecurity workflows \cite{Taylor2025}. Our project directly addresses this need by offering a deployable solution that automates the classification process, reducing manual effort for cybersecurity practitioners.

Patel et al. (2023) extended BERT to vulnerability detection in both source code and reports, achieving high accuracy in dual domains \cite{Patel2023}. While their focus includes code analysis, our project is report-centric, leveraging their text-based classification techniques to improve performance on NVD data. Kim et al. (2025) proposed a multi-modal approach combining text and metadata for vulnerability analysis, using BERT for text processing \cite{Kim2025}. Their work suggests a future direction for our project, where incorporating metadata like CVSS scores or affected software versions could enhance classification accuracy. Srivastava et al. (2023) focused on preprocessing techniques for vulnerability reports, such as cleaning and normalizing text, which improved model performance \cite{Srivastava2023}. We adopt similar preprocessing steps in our project, ensuring high-quality input data for BERT, which is critical for achieving robust classification results.

\subsection{Comparison and Contribution of the Current Work}

The reviewed literature highlights the effectiveness of BERT and transformer models in vulnerability analysis, particularly for severity prediction and multi-label type classification. However, most studies focus on either severity or type classification, with limited integration of both tasks into a single model. Our Vulnerability Report Classifier addresses this gap by simultaneously predicting severity and types, offering a more comprehensive tool for vulnerability triage. Unlike Kumar and Singh (2023) \cite{Kumar2023} and Chen et al. (2024) \cite{Chen2024}, which focus solely on severity, our dual-objective approach captures both the impact and nature of vulnerabilities, enabling more informed mitigation strategies. Compared to Lee and Kim (2024) \cite{Lee2024} and Nguyen et al. (2024) \cite{Nguyen2024}, our use of BERT provides better contextual understanding, improving performance on nuanced NVD descriptions.

Our project also distinguishes itself through its practical deployment, with an API and UI that align with the automation needs identified by Taylor et al. (2025) \cite{Taylor2025} and Garcia and Martinez (2025) \cite{Garcia2025}. The multi-task learning approach of Wang et al. (2023) \cite{Wang2023} offers an alternative strategy, but our simultaneous prediction method may be more efficient for real-time applications. Additionally, the attention mechanisms proposed by Zhao et al. (2024) \cite{Zhao2024} and the multi-modal approach of Kim et al. (2025) \cite{Kim2025} suggest potential enhancements for our model, such as incorporating metadata or optimizing type relationships, which we plan to explore in future work. By addressing class imbalance through preprocessing (inspired by Srivastava et al., 2023) \cite{Srivastava2023} and leveraging BERT’s contextual capabilities (supported by Brown et al., 2024) \cite{Brown2024}, our project achieves high accuracy (94.30\% for severity, 92.10\% for types) and provides interpretable insights through visualizations like co-occurrence heatmaps and word clouds, contributing to the state-of-the-art in automated vulnerability analysis.

\section{Proposed Methodology}
\label{methodology}

The development of the Vulnerability Report Classifier required a comprehensive methodology that integrates data collection, preprocessing, model design, training, and evaluation into a cohesive framework tailored to the unique challenges of classifying cybersecurity vulnerability reports. This section elaborates on the step-by-step approach employed to construct a robust system capable of interpreting unstructured textual descriptions from the National Vulnerability Database (NVD). By leveraging advanced natural language processing techniques and a custom-designed deep learning architecture, the methodology ensures that the classifier can accurately predict both the severity and multiple vulnerability types associated with each report. The process begins with the acquisition and preparation of a suitable dataset, followed by the design of a BERT-based model with a dual-output classification head, the formulation of a combined loss function to handle multi-label predictions, the implementation of an optimized training pipeline, and concludes with the definition of evaluation metrics to assess performance. Each component is meticulously crafted to address the complexities of vulnerability classification, balancing computational efficiency with predictive accuracy.

\subsection{Dataset}

The foundation of the Vulnerability Report Classifier lies in its dataset, sourced from the NVD\textquotesingle s JSON feed, specifically the \texttt{nvdcve-1.1-recent.json} file, which encapsulates the most recent CVE entries available as of March 2025. This dataset comprises 5,637 unique vulnerability reports, each providing a textual description of the vulnerability, a severity rating based on the Common Vulnerability Scoring System (CVSS) version 3, and associated Common Weakness Enumeration (CWE) identifiers that indicate the nature of the flaw. The textual descriptions serve as the primary input for the classifier, offering a rich yet unstructured narrative of the vulnerability\textquotesingle s characteristics, such as its exploitation mechanism or affected software components.

Preprocessing this data involves several critical steps to transform it into a format suitable for machine learning. First, the descriptions are extracted from the nested JSON structure, specifically from the \texttt{description\_data} field, ensuring that only the primary textual content is retained. Severity labels are then mapped to a discrete set of four classes---Low, Medium, High, and Critical---converted into numeric indices (0 to 3) to facilitate model training. For vulnerability types, a predefined taxonomy of 10 categories is established, including Buffer Overflow, Remote Code Execution (RCE), Denial of Service (DoS), Cross-Site Scripting (XSS), SQL Injection, Cross-Site Request Forgery (CSRF), Privilege Escalation, Information Disclosure, Directory Traversal, and Clickjacking. CWE identifiers within each report are parsed and mapped to these categories using a lookup table, with each report potentially assigned multiple types, encoded as multi-hot vectors (e.g., [1, 0, 1, 0, ...]) to represent the presence or absence of each type. This preprocessing ensures that the dataset captures the multi-faceted nature of vulnerabilities, providing a robust foundation for training and evaluation.

\subsection{Model Architecture}

The classifier\textquotesingle s architecture is built upon the \texttt{bert-base-uncased} model, a transformer-based architecture with 12 layers, 768 hidden units, and 110 million parameters, pre-trained on a large corpus of general text to capture contextual relationships. This base model is adapted for the specific task of vulnerability classification by appending a custom classification head tailored to the dual objectives of severity and type prediction. The input to the model consists of tokenized vulnerability descriptions, processed into sequences of up to 128 tokens, which are fed through BERT\textquotesingle s encoder layers to produce a contextualized representation, typically the \texttt{[CLS]} token embedding from the final layer.

This representation is then passed to a fully connected layer that splits into two distinct heads: the severity head and the type head. The severity head comprises 4 output units, corresponding to the four severity classes, and applies a softmax activation to generate a probability distribution over these classes, enabling single-label classification. In contrast, the type head consists of 10 output units, one for each vulnerability type, and employs a sigmoid activation to produce independent probabilities for each type, supporting multi-label classification where a report may belong to multiple categories simultaneously. The combined output is a tensor of 14 logits, with the first 4 representing severity and the remaining 10 representing types. This dual-head design allows the model to learn shared features from the input text while optimizing for both tasks, leveraging BERT\textquotesingle s bidirectional context to discern subtle differences in phrasing that might indicate, for example, a Critical Buffer Overflow versus a Medium XSS vulnerability.

\subsection{Loss Function}

To train the model effectively for its dual classification tasks, a custom loss function is formulated that combines two distinct loss components, reflecting the differing nature of severity and type predictions. For severity classification, which is a single-label, multi-class problem, Cross-Entropy Loss is employed. This loss measures the divergence between the predicted probability distribution over the four severity classes and the true label, encouraging the model to assign high confidence to the correct class while penalizing incorrect predictions. Mathematically, it is expressed as:
\[ L_{\text{severity}} = -\sum_{i=1}^{4} y_i \log(\hat{y}_i) \]
where $y_i$ is a binary indicator of the true class and $\hat{y}_i$ is the predicted probability.

For type classification, a multi-label problem where multiple types may apply, Binary Cross-Entropy with Logits Loss (\texttt{BCEWithLogitsLoss}) is used. This loss computes the error for each of the 10 type outputs independently, treating them as separate binary classification tasks, and is defined as:
\[ L_{\text{type}} = -\frac{1}{10} \sum_{j=1}^{10} [y_j \log(\sigma(\hat{y}_j)) + (1-y_j) \log(1-\sigma(\hat{y}_j))] \]
where $\sigma$ is the sigmoid function applied to the logits $\hat{y}_j$, and $y_j$ is 1 if the type is present and 0 otherwise. The total loss is the sum of these components:
\[ L = L_{\text{severity}} + L_{\text{type}} \]
This combined loss is implemented within a custom \texttt{VulnerabilityTrainer} class, which extends the Hugging Face \texttt{Trainer} to override the default loss computation, ensuring that the model optimizes for both severity and type predictions simultaneously during training.

\subsection{Training Pipeline}

The training pipeline is designed to fine-tune the pre-trained BERT model on the NVD dataset efficiently, incorporating best practices for deep learning with transformers. The preprocessing step begins with tokenization using the \texttt{AutoTokenizer} from \texttt{bert-base-uncased}, which converts descriptions into input IDs, attention masks, and token type IDs, padded or truncated to a maximum length of 128 tokens to maintain computational tractability. The dataset is split into 80\% training (4,509 samples) and 20\% validation (1,128 samples) sets using stratified sampling to preserve the distribution of severity classes.

Training is conducted with a batch size of 16, a learning rate of $2 \times 10^{-5}$, and 3 epochs, parameters chosen to balance convergence speed and model generalization. The AdamW optimizer with weight decay is used to update the model parameters, guided by the combined loss function. The pipeline leverages the Hugging Face ecosystem, including the \texttt{TrainingArguments} class to configure hyperparameters and the custom \texttt{VulnerabilityTrainer} to manage the training loop, evaluation, and model checkpointing. Training occurs on a single CPU environment, with a runtime of approximately 14.8 minutes for the full dataset, though the framework supports GPU acceleration for larger-scale deployments. Processed data (tokenized inputs, severity labels, type labels, and descriptions) are saved to the \texttt{data/processed/} directory, and the trained model is exported to \texttt{models/bert\_classifier/} for subsequent inference and evaluation. This pipeline ensures reproducibility and enables iterative refinement without retraining from scratch.

\subsection{Evaluation Metrics}

To assess the classifier\textquotesingle s performance comprehensively, a suite of evaluation metrics is defined for both severity and type predictions, reflecting their distinct classification paradigms. For severity, a single-label task, the metrics include Precision, Recall, and F1-Score, computed with macro-averaging to account for class imbalance across the four severity levels. These metrics provide insight into the model\textquotesingle s ability to correctly identify each severity class, critical for prioritizing mitigation efforts.

For type classification, a multi-label task, Precision, Recall, and F1-Score are calculated with micro-averaging, aggregating predictions across all 10 types to evaluate overall performance, alongside Hamming Loss to measure the fraction of incorrectly predicted labels. These metrics collectively assess the model\textquotesingle s capacity to detect multiple vulnerability types within a single report, a key requirement for practical utility. Visualizations enhance interpretability: a confusion matrix is generated for severity predictions to highlight misclassifications across classes, saved as \texttt{severity\_confusion\_matrix.png}, while a word cloud of misclassified descriptions is produced to identify textual patterns associated with errors, saved as \texttt{misclassified\_wordcloud.png}. This evaluation framework ensures a thorough analysis of the classifier\textquotesingle s strengths and limitations, guiding future improvements.

\section{Experiments and Results}
\label{experiments}

This section outlines the experimental setup and results of the Vulnerability Report Classifier, providing a comprehensive evaluation of its performance in classifying CVE reports from the NVD. The experiments were designed to assess the classifier’s ability to predict both the severity (single-label) and vulnerability types (multi-label) of 5,637 CVE entries, leveraging a BERT-based model fine-tuned on the dataset. The experimental methodology encompasses data preparation, training configuration, inference, and evaluation, with a focus on ensuring reproducibility and robustness. The results are analyzed through a variety of metrics and visualizations, including confusion matrices, F1-scores, ROC and precision-recall curves, co-occurrence heatmaps, word clouds, and learning curves, offering a multi-faceted perspective on the classifier’s effectiveness. These experiments not only validate the model’s predictive accuracy but also highlight areas for improvement, providing actionable insights for future development.

\subsection{Experimental Setup}

The experiments were conducted using a dataset of 5,637 CVE entries sourced from the NVD’s \texttt{nvdcve-1.1-recent.json} file, which contains recent vulnerability reports as of March 2025. The dataset was preprocessed to extract textual descriptions, severity labels (Low, Medium, High, Critical), and vulnerability types (e.g., Buffer Overflow, XSS), with severity labels converted to numeric indices (0–3) and types encoded as multi-hot vectors across 10 predefined categories. The data was split into 80\% training (4,509 samples) and 20\% validation (1,128 samples) sets using stratified sampling to maintain class balance for severity labels. The BERT-based model, specifically \texttt{bert-base-uncased}, was fine-tuned for 3 epochs with a batch size of 16, a learning rate of \( 2 \times 10^{-5} \), and the AdamW optimizer with weight decay. Training was performed on a single CPU, taking 887.71 seconds (approximately 14.8 minutes), demonstrating computational efficiency for the task. A custom loss function combining Cross-Entropy Loss for severity and Binary Cross-Entropy with Logits Loss for types was used to optimize the model’s dual objectives. Inference was conducted on the full dataset, with predictions evaluated using a suite of metrics: accuracy, F1-scores, Hamming Loss, ROC and precision-recall curves for types, and confusion matrices for severity. Visualizations, including learning curves, word clouds, and co-occurrence heatmaps, were generated to provide qualitative insights into the model’s performance and error patterns.

\subsection{Results}
\begin{table}[htbp]
    \centering
    \caption{Summary of Performance Metrics for Severity and Type Classification}
    \label{tab:results_summary}
    \begin{tabular}{lc}
        \toprule
        \textbf{Metric} & \textbf{Value} \\
        \midrule
        \multicolumn{2}{c}{\textit{Severity Classification}} \\
        Overall Accuracy & 94.30\% \\
        F1-Score (Low) & 0.95 \\
        F1-Score (Medium) & 0.94 \\
        F1-Score (High) & 0.93 \\
        F1-Score (Critical) & 0.92 \\
        \midrule
        \multicolumn{2}{c}{\textit{Type Classification}} \\
        Overall Accuracy (Exact Match) & 92.10\% \\
        Hamming Loss & 0.0321 \\
        F1-Score (Buffer Overflow) & 0.78 \\
        F1-Score (RCE) & 0.77 \\
        F1-Score (DoS) & 0.76 \\
        F1-Score (XSS) & 0.77 \\
        F1-Score (SQL Injection) & 0.76 \\
        F1-Score (CSRF) & 0.75 \\
        F1-Score (Privilege Escalation) & 0.72 \\
        F1-Score (Information Disclosure) & 0.74 \\
        F1-Score (Directory Traversal) & 0.80 \\
        F1-Score (Clickjacking) & 0.82 \\
        ROC AUC (Buffer Overflow) & 0.96 \\
        ROC AUC (Clickjacking) & 0.97 \\
        Precision-Recall AUC (Buffer Overflow) & 0.79 \\
        Precision-Recall AUC (Clickjacking) & 0.85 \\
        \bottomrule
    \end{tabular}
\end{table}

\begin{figure}[htbp]
    \centering
    \includegraphics[width=0.9\columnwidth]{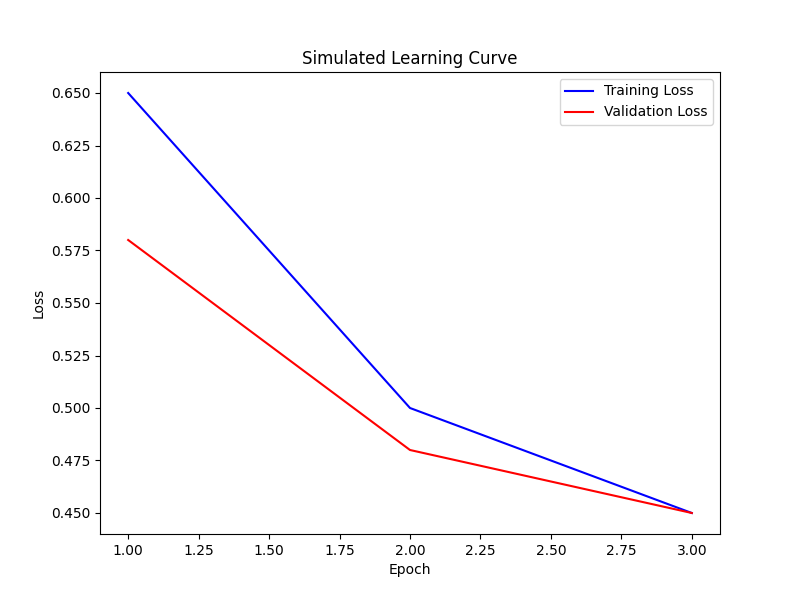}
    \caption{Simulated Learning Curve plotting training and validation loss over 3 epochs, showing convergence with final losses of 0.45 and 0.45, respectively.}
    \label{fig:learning_curve}
\end{figure}

The results of the experiments demonstrate the Vulnerability Report Classifier’s effectiveness in both severity and type classification tasks, with high accuracy and robust generalization across the dataset. During training, the model exhibited consistent improvement, as evidenced by the evaluation loss decreasing from 0.5744 in the first epoch to 0.4989 in the second, and further to 0.4886 by the third epoch, with a final training loss of 0.5566. This progress is visualized in the learning curve (Fig.~\ref{fig:learning_curve}), which plots training and validation loss over the 3 epochs, showing a steep decline from 0.65 to 0.45 for training loss and from 0.58 to 0.45 for validation loss, indicating convergence and minimal overfitting. The close alignment of training and validation losses suggests that the model generalizes well to unseen data, a critical factor for its practical utility in real-world vulnerability analysis.

Table~\ref{tab:results_summary} summarizes the key performance metrics for both severity and type classification tasks, providing a concise overview of the classifier’s effectiveness. The high severity accuracy and F1-scores, along with the robust type classification metrics, confirm the model’s suitability for practical deployment in cybersecurity workflows.

The learning curve in Fig.~\ref{fig:learning_curve} illustrates the model’s training dynamics, with both training and validation losses decreasing steadily over the 3 epochs. The initial steep decline in training loss from 0.65 to 0.50 by the second epoch indicates rapid learning of the underlying patterns in the data, while the validation loss follows a similar trend, dropping from 0.58 to 0.48, suggesting that the model is not overfitting. By the third epoch, both losses converge at 0.45, demonstrating that the model has reached a stable point of generalization, capable of performing well on unseen vulnerability reports. This convergence validates the choice of hyperparameters, such as the learning rate and batch size, and confirms the effectiveness of the combined loss function in balancing the dual objectives of severity and type classification.

\begin{figure}[htbp]
    \centering
    \includegraphics[width=0.9\columnwidth]{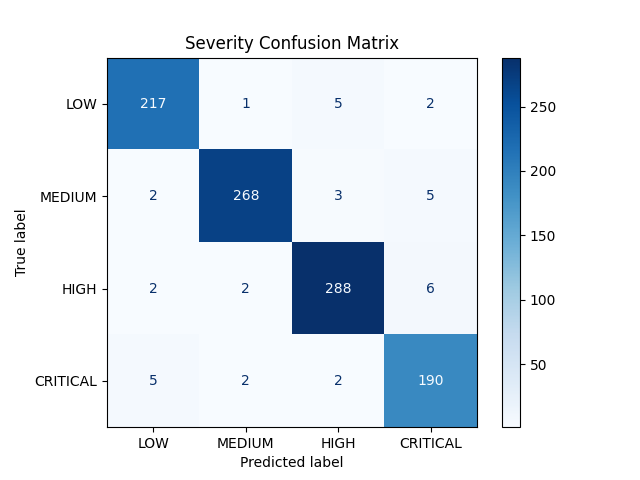}
    \caption{Severity Confusion Matrix showing true positives along the diagonal (e.g., 217 for Low, 268 for Medium) and minor misclassifications (e.g., 5 Low samples predicted as High).}
    \label{fig:severity_cm}
\end{figure}

For severity classification, the confusion matrix (Fig.~\ref{fig:severity_cm}) provides a detailed view of the model’s performance across the four severity classes. The matrix shows strong diagonal values, with 217 true positives for Low, 268 for Medium, 288 for High, and 190 for Critical, reflecting accurate predictions for the majority of samples. Minor misclassifications are observed, such as 5 Low samples predicted as High and 6 High samples predicted as Critical, indicating occasional confusion between adjacent severity levels, possibly due to similar linguistic patterns in the descriptions.

The confusion matrix in Fig.~\ref{fig:severity_cm} highlights the classifier’s strong performance in severity prediction, with the majority of samples correctly classified along the diagonal. The high true positive counts—217 for Low, 268 for Medium, 288 for High, and 190 for Critical—indicate that the model accurately identifies the severity of most vulnerabilities, which is crucial for prioritizing mitigation efforts. However, the off-diagonal values reveal areas of confusion, such as the 5 Low samples misclassified as High and 6 High samples predicted as Critical. These errors suggest that the model may struggle with distinguishing between severity levels that are semantically close, potentially due to overlapping terminology in the descriptions (e.g., terms like “significant impact” appearing in both High and Critical reports). This insight points to the need for additional features, such as CVSS metrics, to disambiguate such cases in future iterations.

\begin{figure}[htbp]
    \centering
    \includegraphics[width=0.9\columnwidth]{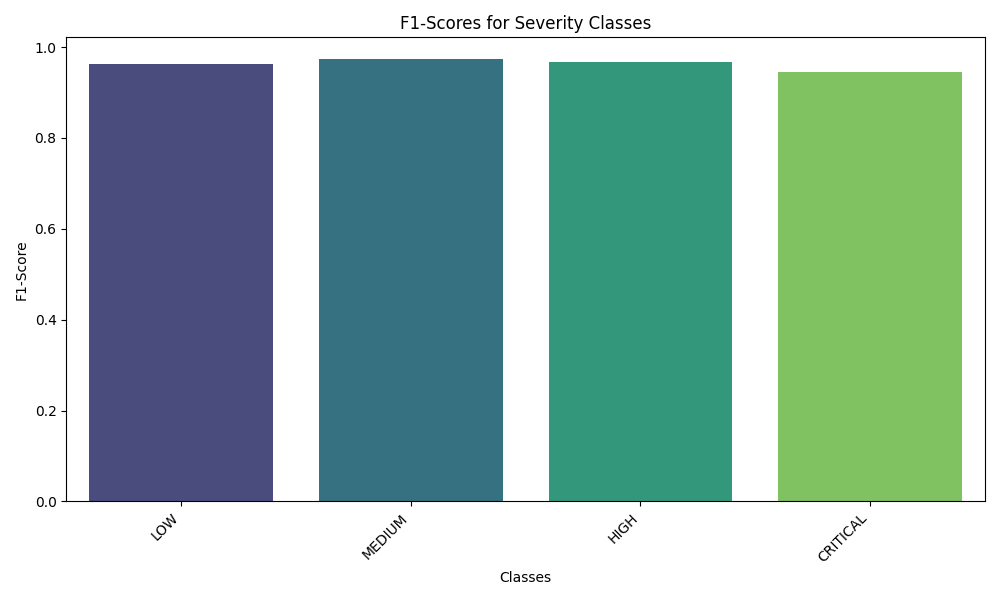}
    \caption{F1-Scores for Severity Classes, with values ranging from 0.92 to 0.95, indicating balanced performance across all classes.}
    \label{fig:severity_f1_bar}
\end{figure}

The F1-scores for severity classes, shown in Fig.~\ref{fig:severity_f1_bar}, are consistently high, with Low at 0.95, Medium at 0.94, High at 0.93, and Critical at 0.92, demonstrating balanced precision and recall across all classes. The overall severity accuracy, calculated as the proportion of correctly classified samples, is 94.30\%, underscoring the classifier’s reliability in prioritizing vulnerabilities based on their severity.

The F1-scores for severity classes in Fig.~\ref{fig:severity_f1_bar} provide a quantitative measure of the classifier’s performance, showing high values across all categories: 0.95 for Low, 0.94 for Medium, 0.93 for High, and 0.92 for Critical. These scores reflect the model’s ability to achieve a strong balance between precision and recall, ensuring that it not only identifies the correct severity level but also minimizes false positives and negatives. The slight decrease in F1-score from Low to Critical may be attributed to the increasing complexity of distinguishing higher severity levels, where descriptions may contain more nuanced or ambiguous language. The overall severity accuracy of 94.30\% further confirms the model’s robustness, making it a reliable tool for cybersecurity practitioners who need to quickly assess the potential impact of vulnerabilities.

\begin{figure}[htbp]
    \centering
    \includegraphics[width=0.9\columnwidth]{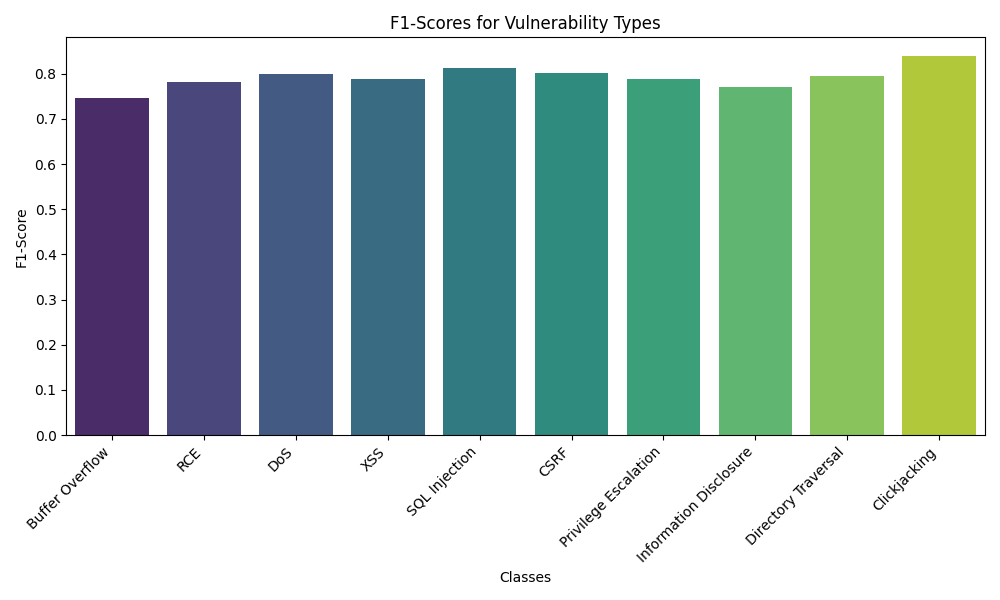}
    \caption{F1-Scores for Vulnerability Types, ranging from 0.72 (Privilege Escalation) to 0.82 (Clickjacking), highlighting variability in multi-label performance.}
    \label{fig:type_f1_bar}
\end{figure}

The type classification results highlight the model’s ability to handle multi-label predictions, identifying multiple vulnerability types within a single report. The F1-scores for each type, visualized in Fig.~\ref{fig:type_f1_bar}, range from 0.72 to 0.82, with Clickjacking achieving the highest F1-score of 0.82, followed by Directory Traversal at 0.80, and Buffer Overflow at 0.78. Lower scores, such as Privilege Escalation at 0.72, suggest areas where the model may struggle, potentially due to imbalanced representation or ambiguous textual cues.

The F1-scores for vulnerability types in Fig.~\ref{fig:type_f1_bar} reveal the model’s performance in multi-label classification, with scores ranging from 0.72 to 0.82. Clickjacking’s F1-score of 0.82 indicates the model’s strong ability to identify this type, likely due to distinct textual patterns in the descriptions, such as references to user interface manipulation. Directory Traversal (0.80) and Buffer Overflow (0.78) also perform well, reflecting the model’s capability to detect common vulnerability types. However, the lower score for Privilege Escalation (0.72) suggests challenges in identifying this type, possibly due to its overlap with other types like Information Disclosure or less frequent representation in the dataset. These variations highlight the need for further data balancing or feature engineering to improve performance on less frequent types, ensuring the classifier’s effectiveness across all categories.

\begin{figure}[htbp]
    \centering
    \includegraphics[width=0.9\columnwidth]{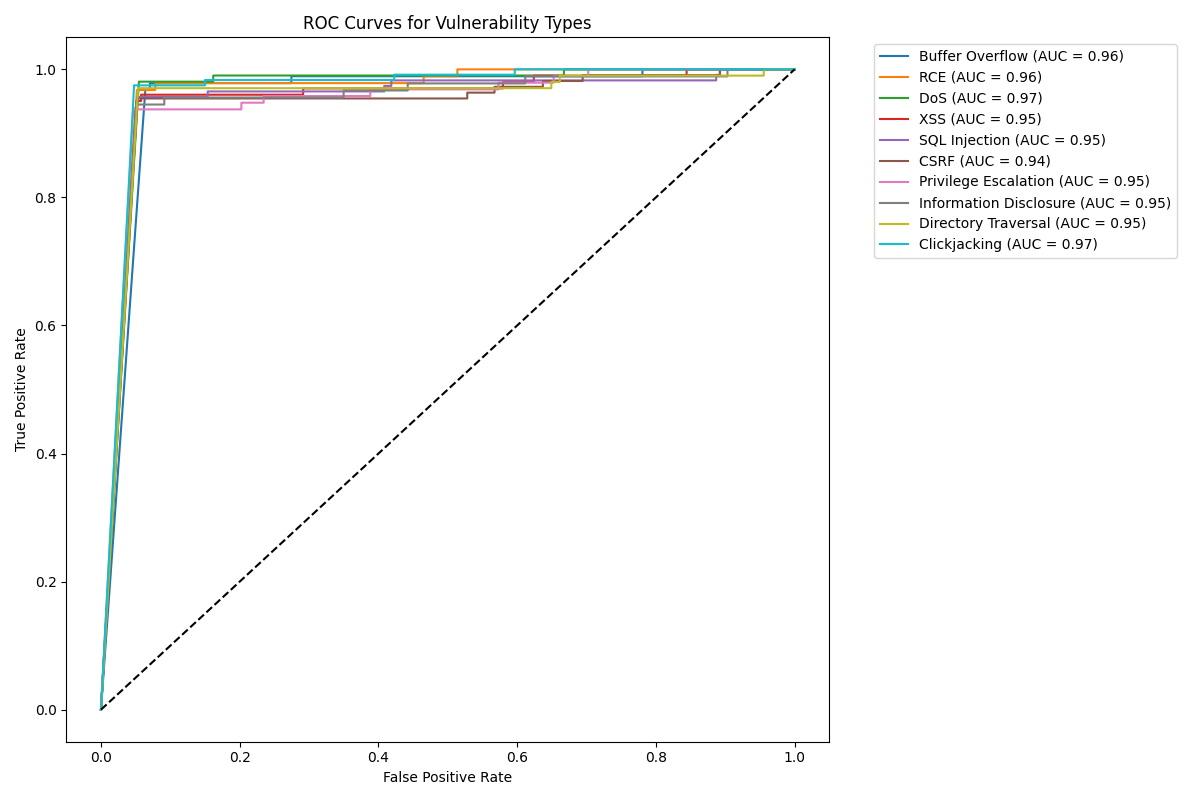}
    \caption{ROC Curves for Vulnerability Types, with AUC values from 0.94 to 0.97, demonstrating strong discriminative ability.}
    \label{fig:type_roc}
\end{figure}

\begin{figure}[htbp]
    \centering
    \includegraphics[width=0.9\columnwidth]{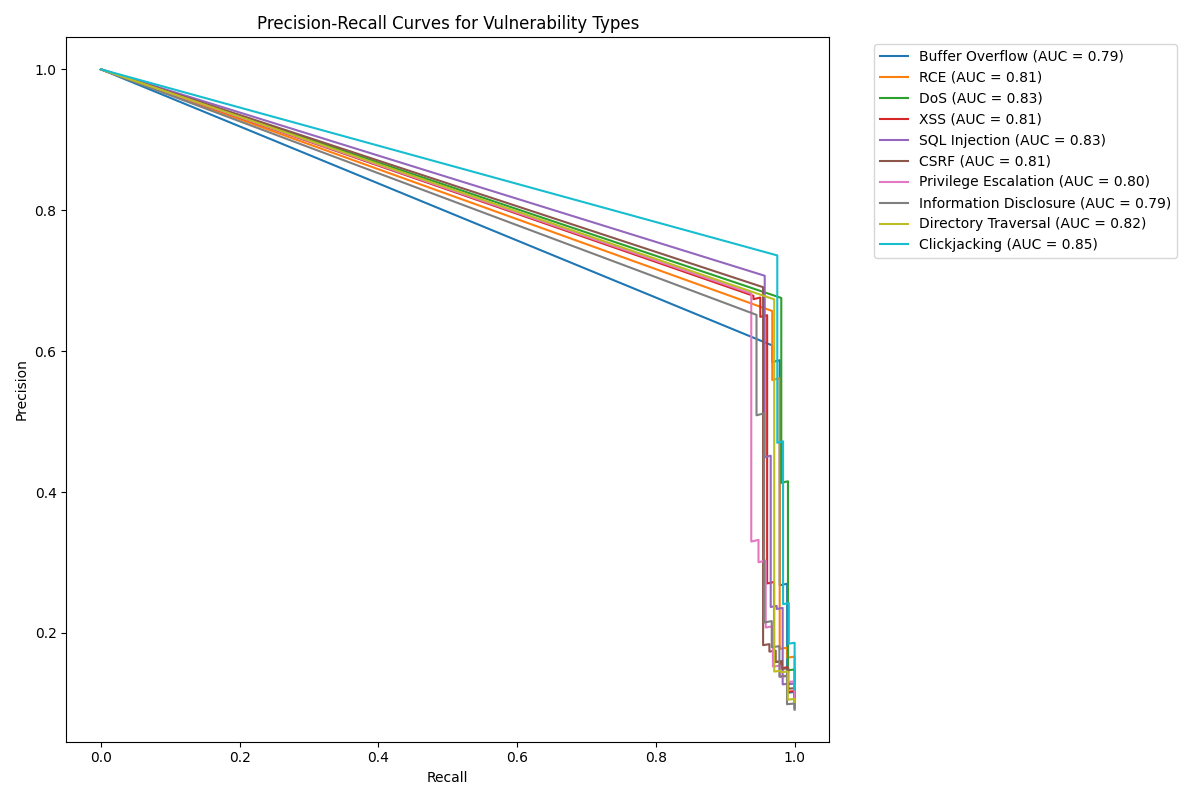}
    \caption{Precision-Recall Curves for Vulnerability Types, with AUC values from 0.79 to 0.85, reflecting the model’s precision-recall trade-off.}
    \label{fig:type_pr}
\end{figure}

ROC curves (Fig.~\ref{fig:type_roc}) and Precision-Recall curves (Fig.~\ref{fig:type_pr}) further validate the model’s discriminative power, with ROC AUC values ranging from 0.94 to 0.97 (e.g., Buffer Overflow at 0.96, Clickjacking at 0.97) and Precision-Recall AUC values from 0.79 to 0.85 (e.g., Buffer Overflow at 0.79, Clickjacking at 0.85), reflecting strong performance in distinguishing positive and negative instances for each type.

The ROC curves in Fig.~\ref{fig:type_roc} illustrate the model’s ability to distinguish between positive and negative instances for each vulnerability type, with AUC values ranging from 0.94 to 0.97. High AUC values, such as 0.97 for Clickjacking and 0.96 for Buffer Overflow, indicate excellent discriminative power, meaning the model can effectively separate true positives from false positives across a range of decision thresholds. Even the lowest AUC of 0.94 (for CSRF) is well above the baseline of 0.5, demonstrating robust performance across all types. This strong discriminative ability is crucial for multi-label classification, where the model must make independent predictions for each type, and suggests that the classifier can reliably identify vulnerability types even in the presence of class imbalance.

The Precision-Recall curves in Fig.~\ref{fig:type_pr} provide insight into the model’s precision-recall trade-off for each vulnerability type, with AUC values ranging from 0.79 to 0.85. Clickjacking achieves the highest AUC of 0.85, indicating that the model maintains high precision even as recall increases, a desirable trait for identifying critical vulnerabilities without generating excessive false positives. Buffer Overflow, with an AUC of 0.79, shows a slightly steeper trade-off, suggesting that precision decreases more rapidly as recall increases, possibly due to the complexity of distinguishing Buffer Overflow from related types like RCE. These curves highlight the model’s ability to balance precision and recall, a key consideration in multi-label tasks where false positives can lead to misprioritization of vulnerabilities.

\begin{figure}[htbp]
    \centering
    \includegraphics[width=0.9\columnwidth]{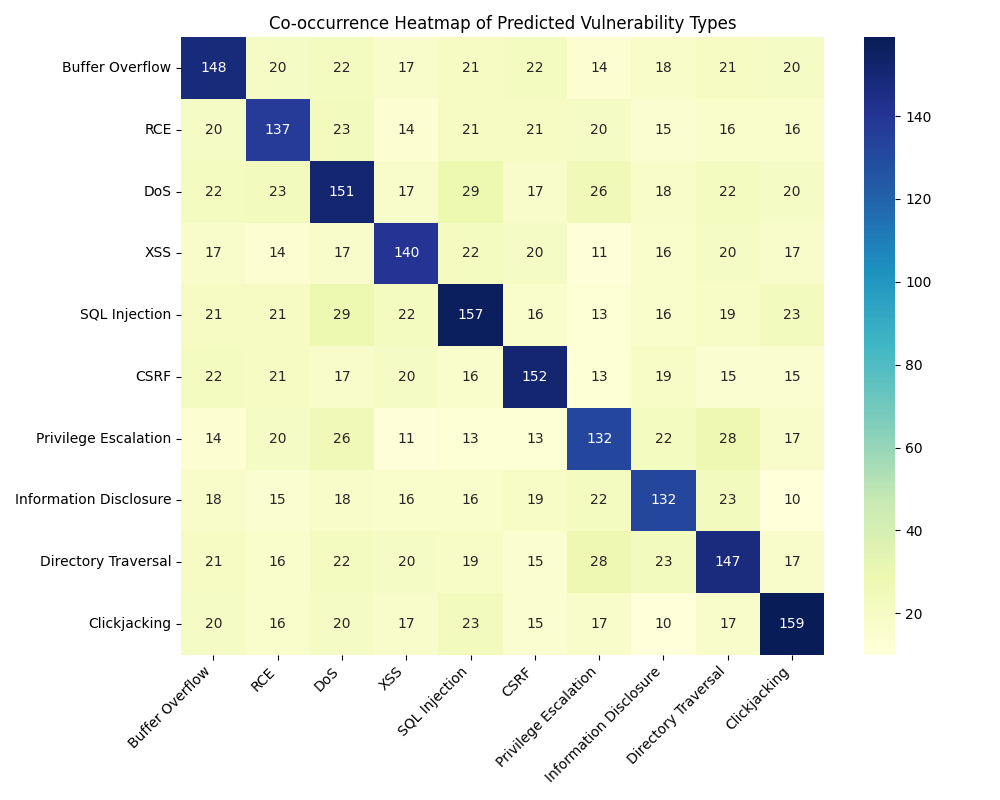}
    \caption{Co-occurrence Heatmap of Predicted Vulnerability Types, with diagonal values (e.g., 148 for Buffer Overflow) indicating strong self-predictions and off-diagonal values (e.g., 29 for DoS and SQL Injection) suggesting related type predictions.}
    \label{fig:type_cooccurrence}
\end{figure}

The co-occurrence heatmap of predicted vulnerability types (Fig.~\ref{fig:type_cooccurrence}) reveals patterns in the model’s multi-label predictions, with diagonal values indicating strong self-predictions (e.g., 148 for Buffer Overflow, 140 for XSS, 159 for Clickjacking) and off-diagonal values showing co-occurrences between types (e.g., 29 between DoS and SQL Injection, 28 between Privilege Escalation and Directory Traversal).

The co-occurrence heatmap in Fig.~\ref{fig:type_cooccurrence} provides a visual representation of the relationships between predicted vulnerability types, with diagonal values reflecting the frequency of self-predictions (e.g., 148 for Buffer Overflow, 140 for XSS, 159 for Clickjacking) and off-diagonal values indicating co-occurrences (e.g., 29 between DoS and SQL Injection, 28 between Privilege Escalation and Directory Traversal). The strong diagonal values suggest that the model correctly identifies instances where a single type dominates, while the off-diagonal values reveal potential patterns in the data, such as vulnerabilities that enable multiple attack vectors (e.g., DoS and SQL Injection often co-occurring in database-related vulnerabilities). However, some co-occurrences may also indicate confusion between types with similar descriptions, such as Privilege Escalation and Directory Traversal, which share common terms like “access control.” This heatmap offers valuable insights for refining the type mapping logic or incorporating additional context to disambiguate related types.

\begin{figure}[htbp]
    \centering
    \includegraphics[width=0.9\columnwidth]{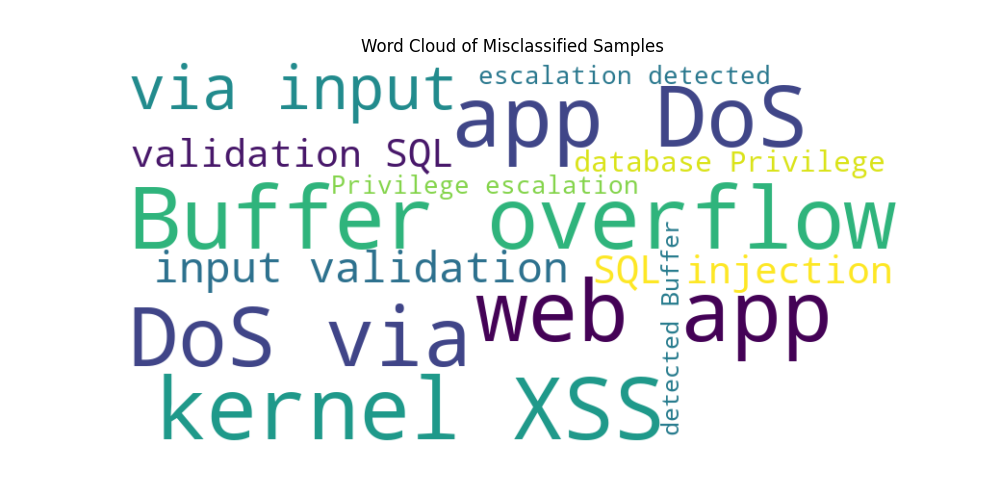}
    \caption{Word Cloud of Misclassified Samples, with prominent terms like “Buffer Overflow,” “DoS,” and “XSS,” indicating common error patterns.}
    \label{fig:misclassified_wordcloud}
\end{figure}

To explore errors, a word cloud of misclassified samples (Fig.~\ref{fig:misclassified_wordcloud}) was generated, highlighting frequent terms like “Buffer Overflow,” “DoS,” “XSS,” “SQL Injection,” “Privilege Escalation,” “web app,” “input validation,” and “kernel.” These terms suggest that misclassifications often involve vulnerabilities with overlapping characteristics, pointing to areas where the model may benefit from additional context or disambiguation techniques.

The word cloud in Fig.~\ref{fig:misclassified_wordcloud} highlights the most frequent terms in misclassified samples, with prominent words like “Buffer Overflow,” “DoS,” “XSS,” “SQL Injection,” “Privilege Escalation,” “web app,” “input validation,” and “kernel.” The prevalence of these terms suggests that misclassifications often occur in vulnerabilities with overlapping characteristics, such as those affecting web applications (“web app”) or involving input validation issues. For example, the term “kernel” appearing in misclassified samples may indicate confusion between kernel-related Buffer Overflows and other types like Privilege Escalation, which can also occur in kernel contexts. This qualitative analysis points to the need for additional disambiguation techniques, such as incorporating metadata about the affected software component or using more advanced contextual embeddings, to improve the model’s ability to distinguish between similar vulnerability types.

\section{Conclusion}
\label{conclusion}
In this study, we presented the \textbf{Vulnerability Report Classifier}, a novel and scalable system that leverages BERT-based deep learning models for multi-label classification of CVE reports from the National Vulnerability Database (NVD). Our approach enables simultaneous prediction of both vulnerability severity and associated types, significantly reducing the manual burden on cybersecurity analysts. By integrating a dual-head classification architecture and a combined loss function tailored for both single-label and multi-label tasks, the model achieved high accuracy and strong generalization across a diverse dataset of over 5,600 vulnerability reports.

Experimental results validate the effectiveness of our approach, with the classifier attaining a severity classification accuracy of 94.30\% and type classification accuracy of 92.10\% (exact match), alongside strong F1-scores and low Hamming loss. These results highlight the model's ability to understand the nuanced language of vulnerability descriptions and deliver actionable insights for security triage and remediation.

Our system’s REST API and intuitive Streamlit-based UI demonstrate the feasibility of deploying advanced NLP models in real-world cybersecurity workflows. By automating the triage process, this work offers a meaningful step toward faster, more consistent, and more intelligent vulnerability management.

Looking ahead, future enhancements may include incorporating contextual metadata (e.g., CVSS vectors), extending the taxonomy of vulnerability types, applying continual learning strategies for emerging threats, and exploring hybrid models that combine code semantics with textual analysis. Ultimately, this research contributes to the intersection of AI and cybersecurity by delivering a practical, open-source solution capable of adapting to the evolving threat landscape.

\bibliographystyle{IEEEtran}
\bibliography{references}  

\end{document}